\documentclass[10pt,superscriptaddress,twocolumn,showkeys,aps,prb]{revtex4-1}
 
\usepackage{graphicx}
\usepackage{dcolumn}
\usepackage{bm}
\usepackage[latin1]{inputenc}
\usepackage{float}
\usepackage{color}
\usepackage{amsmath}
\usepackage{booktabs}

\definecolor{rojo}{rgb}{1,0,0}
\definecolor{verde}{rgb}{0,0.8,0.2}
\definecolor{azul}{rgb}{0,0,1}
\definecolor{rosa}{cmyk}{0,1,0,0}

\usepackage{soul,xcolor}

\usepackage{latexsym}
\usepackage{amsmath}
\usepackage{bm}
\usepackage{amsthm}
\usepackage{bbm}
\usepackage{amsfonts}
\usepackage{amssymb}
\usepackage{color}
\usepackage{epsfig}
\usepackage{hyperref}
\hypersetup{colorlinks=true, citecolor=blue}
\hypersetup{colorlinks=true, citecolor=blue}
\usepackage{soul}

\setstcolor{red}

\begin{document}

\title{ Induced exchange and spin-orbit effects by proximity in graphene on Ni and Co }

\author{Mayra Peralta}
\email{mperalta@yachaytech.edu.ec}
\affiliation{Departamento de F\'isica - Centro de Nanociencias y Nanotecnolog\'ia, Universidad Nacional Aut\'onoma de M\'exico, Apdo. Postal 14, 22800 Ensenada, Baja California, M\'exico}
\affiliation{Yachay Tech University, School of Physical Sciences \& Nanotechnology, 100119-Urcuqu\'i, Ecuador}

\author{Ernesto Medina}
\affiliation{Yachay Tech University, School of Physical Sciences \& Nanotechnology, 100119-Urcuqu\'i, Ecuador}
\affiliation{Centro de F\'isica, Instituto Venezolano de Investigaciones Cient\'ificas, 21827, Caracas 1020 A, Venezuela.}

\author{Francisco Mireles}
\email{fmireles@cnyn.unam.mx}
\affiliation{Departamento de F\'isica - Centro de Nanociencias y Nanotecnolog\'ia, Universidad Nacional Aut\'onoma de M\'exico, Apdo. Postal 14, 22800 Ensenada, Baja California, M\'exico}

\date{\today}

\begin{abstract}
 
The induced-proximity effects of nearly commensurate lattice structure of a graphene layer on Ni(111) and Co(0001) substrates in the  AC stacking configuration are addressed through an analytical tight-binding approach within the Slater-Koster method. A minimal Hamiltonian is constructed by considering the hybridizations of the magnetic $3d$-orbitals of Ni(Co) atoms with the $p_z$-orbitals of graphene, in addition to the atomic spin-orbit coupling and the magnetization of the Ni(Co) atoms.  A low-energy effective Hamiltonian for graphene/Ni(Co) describing  the perturbed $\pi$-bands in the vicinity of the Dirac points is derived which enable us to get further insight on the physical nature of the induced-effective couplings to the graphene layer. It is shown that a magneto-spin-orbit type effect may emerge through two competing mechanisms simultaneously present, namely the proximity induced exchange and Rashba spin-orbit interaction.  Such effects  results  in giant exchange splittings and robust Rashba spin-orbit coupling transferred  to the graphene layer in agreement with recent density functional theory calculations and experimental observations. We further analyze the physical conditions for the appearance of intact Dirac cones in the minority spin bands as observed by recent photoemission measurements with spin resolution.

\end{abstract}

\maketitle

\section{Introduction}

Graphene in proximity with metallic substrates may acquire fascinating new electronic and magnetic properties. \cite{Novoselov,Voloshina}
Depending upon the appropriate substrate, a number of interesting features may arise in its band structure, such as the appearance of an energy gap between the $\pi$ (electron) and $\pi^*$ (hole) bands, the shifting of the Fermi energy $E_F$ of the Dirac cone, distortions of its linear dispersion, and the formation of local band gaps. \cite{Giovannnetti,Wintterlin, Klimo,Khomyakov, DFTReference} The main known mechanisms responsible for these band modifications  are in one hand, the strong hybridization between the $2p_z$ orbitals of the carbon atoms with the $d$-states of the metallic layer, and on the other, the sublattice symmetry breaking. Remarkably, these proximity effects may also promote the emergence of the anomalous Hall effect in graphene/ferromagnetic films,\cite{WangProximity} quantum spin-Hall phases in graphene/topological-insulating interfaces,\cite{KHJin} and the enhancement of the Rashba spin-orbit interaction.\cite{Marchenko2012, Varykhalov1,Rashba, Krivenkov, Otrokov, Cysne, Mayra_etal}

The topic of induced spin-orbit effects in graphene by proximity is of especial prominence for spintronics applications given the smallness of intrinsic spin-orbit splitting effects in freestanding graphene ($\lesssim 50$ $\mu$eV).\cite{Konschuh, Min}  There is a strong experimental evidence that the epitaxial synthesis of graphene interfaces with high spin-orbit metals ({\it e.g.} Au, Ag, Pb) intercalated or in direct contact with different substrates such as Ir, Ni, and Co, leads to unusually large Rashba spin-orbit splittings of the Dirac cones that ranges between 13-100 meVs. \cite{Marchenko2012, Varykhalov1,Rashba, Krivenkov, Otrokov,Zhizhin}  It is also established that the proximity spin-orbit coupling increases with the atomic number of the transition metal.  As supported by  several density functional theory (DFT) studies, \cite{DFTReference,Marchenko2012,Krivenkov,Zhizhin}  the rather anomalous increase of the spin-splitting is ascribed to a strong  $\pi$-$d$ hybridization between graphene and the substrate and/or the intercalated layer.
  
The direct proximity with ferromagnetic substrates, on the other hand, leads to a natural breaking of time reversal symmetry and to the transfer of exchange fields to graphene.\cite{Macdonald2013,Yang, Peralta2016,Vo} This exchange coupling in conjunction with spin-orbit interaction opens the possibility of a wider spin-dependent extension of graphene outstanding characteristics.\cite{Phong}  It is important to remark that in the process of transferring ferromagnetic properties to graphene, it is also highly desirable to preserve its attractive features of robust mobility and linearity of its electronic bands near the Fermi level.\cite{Usachov,Marchenko2012} These physical conditions  have proved however, to be extremely challenging to realize experimentally. For instance, recent experiments\cite{Usachov,Marchenko2012,Macdonald2013} have reported  that the synthesis of graphene on ferromagnetic Ni(111) and Co(0001) in the AC stacking configuration, leads to strong exchange splittings with the preservation of the linearity of the bands at the Dirac point, however no Rashba coupling splitting was observed in these experiments. 

More recent angle- and spin-resolved photoemission experiments (SARPES) by Rybkin et al. \cite{Rybkin} investigated the spin-dependent band splitting in graphene/Co(0001)  intercalated with Au. Their measurements suggests that the combined action of a strong exchange and Rashba coupling in quasi-freestanding graphene is indeed possible. From the analysis of their spin-resolved photoemission spectra and DFT calculations the authors were able to determine the existence of a strong Rashba spin-orbit splitting of around $57$ meV (most likely due to the Au atoms) and a giant exchange splitting of the Dirac cone of $\sim 175$ meV; all this whiles the linear dispersion was nevertheless largely preserved comprising evidence of a so called magneto-spin-orbit effect.\cite{Rybkin} However, the presence of Co atoms into the intercalated Au layer was not fully excluded in their XPS analysis. Clearly there is a need of a deeper understanding  at the microscopical level of the physical nature of the proximity induced Rashba and exchange couplings,  its magnitude estimates, as well as its interplay in graphene/ferromagnetic interfaces. 
Having a better understanding of these coupling mechanisms could also serve as an excellent platform to investigate, interesting topological phenomena and the realization of the quantum anomalous Hall effect,   in addition to its relevance for applied studies in new spintronic devices.
 
Motivated by these works, the aim of this paper is to introduce a simplified microscopic theoretical model of the effective interactions in a AC stacking graphene/Ni and graphene/Co interfaces based in a nearest-neighbor multiorbital tight-binding approach.  The model will allow us to gain further insight to the origins of the observed large Rashba spin-orbit and exchange couplings, proximity-induced at the graphene layer, and its relative estimates.  Although our primary interest here is on Ni and Co substrates, our results are applicable to a broad range of {\it 3d}-ferromagnetic metals, as we expect that the overall qualitative features of the proximity induced exchange and spin-orbit coupling effects shall hold as well. The study permits to explore on the microscopic physical conditions that leads to the observed preservation of the linearity of the bands despite the strong  $\pi$-$d$ hybridization in such interfaces. In particular,  we derive an effective low energy Hamiltonian for graphene/Ni and graphene/Co in the vicinity of the $K(K')$ Dirac points that captures the main characteristics of the bands reported experimentally and by DFT studies for the case of graphene/Ni and graphene/Co \cite{Varykhalov, Marchenko2012,Usachov} yielding as well a qualitative resemblance to the observed results of a magneto-spin-orbit effect in graphene/Au/Co interfaces. \cite{Marchenko2015,Rybkin} The model helps us to determine which 3$d$-orbitals of the ferromagnetic layer participate predominantly in the strong hybridizations with the $\pi$-bands and originates the linear electronic structure in the vicinity of the Dirac point. We also analyze the appearance of intact (unperturbed) Dirac cones in the minority spin bands observed by several angle- and spin-resolved photoemission experiments.\cite{Varykhalov, Marchenko2015, Usachov}

\section{MODEL}
\subsection{Graphene/N\lowercase{i}(C\lowercase{o}) Interface}
\begin{figure}
\begin{center}
\includegraphics[width=6.0cm,height=8.0cm]{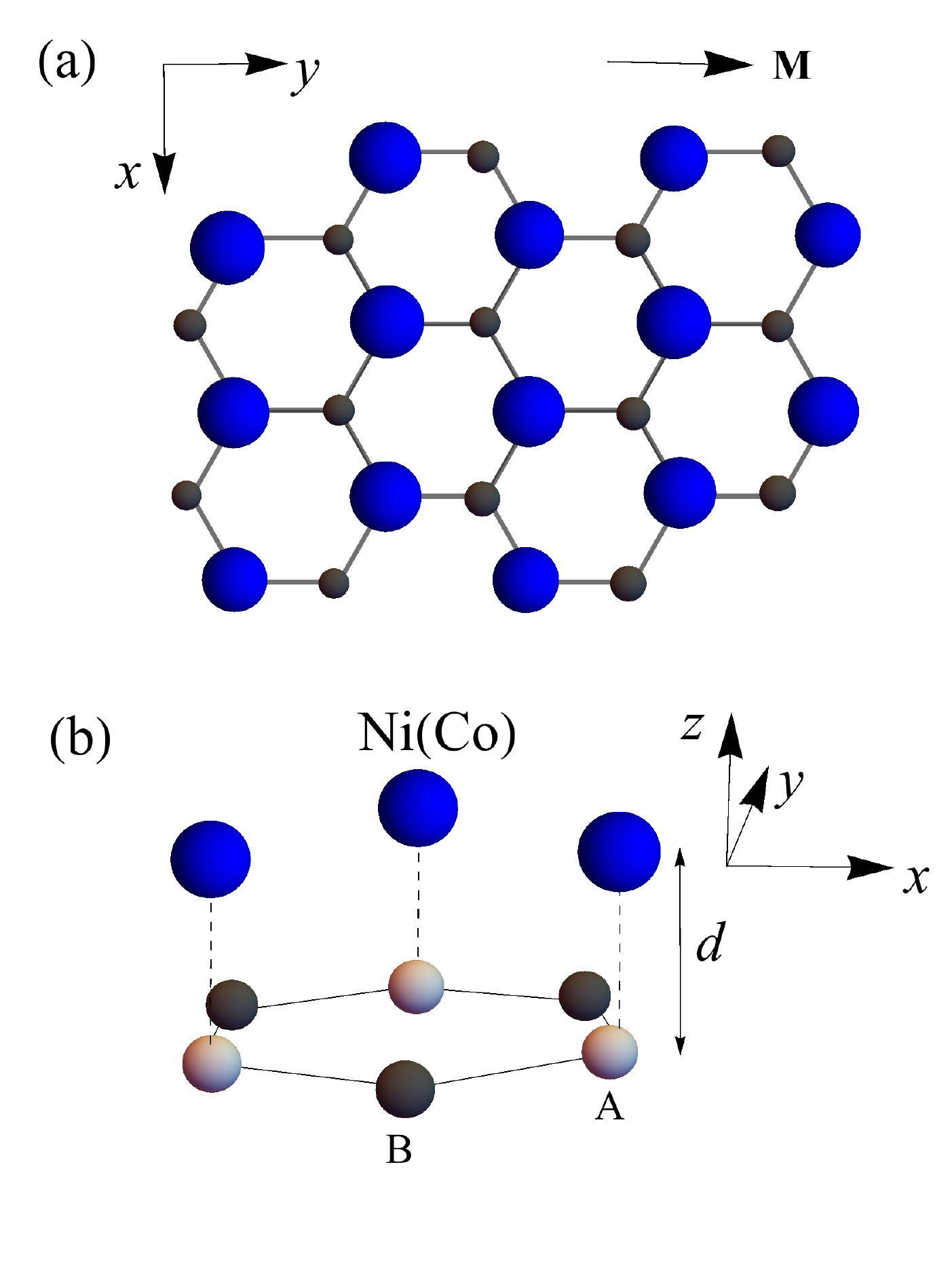}
\end{center}
\vspace{-1.2cm}
\caption{\small Schematic of the graphene/Ni(Co) under study. (a): Top view of the configuration studied (AC stacking), where Ni(Co) atoms are directly over the atoms of sublattice A and atoms of the sublattice B are in the hcp sites of the Ni(Co) lattice. The direction of the Ni(Co) magnetic moment $\mathbf{M}=M\hat{y}$ is shown. (b): 3D view of a single graphene hexagon which shows the Ni(Co)-graphene distance $d$.}
\label{system}
\end{figure}
\vspace{-0.5cm}

Consider a graphene monolayer deposited on a magnetized nickel(cobalt) substrate in which the arrangements of magnetic atoms form a commensurate lattice with negligible mismatch. The layered system is supposed to be strainless and without disorder.  In order to study the interactions between graphene and a Ni(Co) substrate we will consider only the Ni(Co) upper layer in contact with graphene, as the magnetic couplings are expected to decay exponentially with the distance. It has been reported that the the most stable vertical arrangement of the atoms \cite{Macdonald2013, Varykhalov}  is the AC stacking configuration for both substrates, in which the Ni(Co) atoms are placed directly over the atoms of sublattice A of graphene and atoms of the sublattice B are in the hcp sites of the Ni(Co) lattice,  Fig.~\ref{system}(a).  The equilibrium Ni-graphene distance reported in the literature\cite{Krivenkov} for this configuration  is of $d=2.05$\AA, while the similar distance between Co-graphene is reported\cite{Varykhalov} to be $2.11$\AA (see Fig.~\ref{system}(b)). In addition we will assume that the magnetic coupling between nickel(cobalt) atoms is negligible. Henceforth, as indicated in Fig.~\ref{system}(a), the direction of the magnetization (magnetic moment), $\mathbf{M}$, for the Ni(Co) layer is chosen along the $y$-axis.

\subsection{Tight-binding model  and Low Energy Effective Hamiltonian}

The simplest nearest-neighbor tight-binding model that describes graphene in contact with a ferromagnet metal interface in the AC-stacking configuration shall takes into account the carbon valence $2p_z$ orbitals that form the electron $\pi $ and hole $\pi^*$ bands in pristine graphene, and the outer shell orbitals of nickel(cobalt) atoms, namely $3d_{xy}$, $3d_{xz}$, $3d_{yz}$, $3d_{3z^2-r^2}$,  and $3d_{x^2-y^2}$ orbitals. \cite{Konschuh, Min}  We consider a two center tight-binding Hamiltonian that incorporates the atomic spin-orbit interaction and the exchange field due the nearby nickel(cobalt) atoms.  A Slater-Koster Hamiltonian that dictates the main hybridization properties is constructed with matrix elements of the form $\langle\phi_{\mu,s}|\hat{H}_{_{TB}}|\phi_{\mu',s'} \rangle$ where $\mu=\{0,1,2,3,4,5,6\}$ denotes the $2p^{A/B}_z$ orbitals of carbon atoms and the five $3d$-orbitals of nickel(cobalt), and $s=\{\uparrow,\downarrow\}$ labels the electron spin, yielding a $14\times14$ overlapping matrix for the graphene/Ni(Co) interface (Appendix A). A low energy $4\times4$ effective Hamiltonian in the momentum space is then derived and it has the compact form at the $K(K')$ valley 
\begin{equation}
{\cal H}_{{\it eff}}(\bm{p})  = (\mathbf{\bm \sigma}\, \cdot\, \bm{p})\otimes (v^{*}_{F}s_{o} - v^{*}_{d}s_{y}) + \lambda_{_{\it XR}}(\mathbf{\bm\sigma}{\bm \otimes}\bm{s})_{z} + {  \cal E}_{{ex}},  \\
\label{heffpz1}
\end{equation}

\noindent where ${\bm \sigma}=$ $ (\xi\sigma_x,\sigma_y)$ is the Pauli vector matrix acting on the sublattice \{$A,B$\} in pristine graphene, with $\xi= +1(-1)$ for the $K(K')$ valley, respectively,  $\bm p$ is the momentum operator, ${{\bm s}=(s_x,s_y,s_z)}$ is the Pauli vector matrix that acts on the real spin, with $s_o$ the unit matrix in spin-space.  
In Eq.~(\ref{heffpz1}) the first term to the right corresponds to an effective Dirac term linear in momentum with a velocity $v^{*}_F$ diagonal in spin-space shifted by an off-diagonal exchange field dependent velocity term $v^{*}_d$. The second term has the form of a Rashba spin-orbit interaction $(\mathbf{\bm\sigma}{\bm \otimes}\bm{s})_{z}\equiv\xi\sigma_{x}\otimes s_y -\sigma_{y}\otimes s_x $, with a coupling parameter  $ \lambda_{_{\it XR}}$. The proximity-induced Rashba term is present due the hybridization of the $p_z$-orbitals with the $d_{3z^2-r^2}$ as well as for the presence to the on-site exchange field of nickel(cobalt) atoms. Lastly the term ${  \cal E}_{_{ex}}$ represent a constant matrix which effectively induces 
an exchange-dependent energy shifting of the Dirac cones together with a exchange-field transfer to the graphene layer along the magnetization orientation of the ferromagnetic substrate.  As we shall discuss, the interplay of the last two terms in Eq.~(\ref{heffpz1}) accounts for the magneto-spin-orbit effect. The model effective Hamiltonian described in Eq.~(\ref{heffpz1}) captures very well the main features of the observed low energy bands and to first principles calculations of graphene on nickel(cobalt), as discussed in detail in the next sections.

We find that the reduced velocity $ v^{*}_{_F} =  v_{_F} +\eta_{_d}\Delta_0$ and $v^{*}_{d}=\eta_{d}\Delta_{ex}$,  where  $v_{_F}=-\frac{\sqrt{3}a}{2\hbar}V_{pp\pi}$  is the Fermi velocity for freestanding graphene in terms of the relevant Slater-Koster overlapping integral between carbon-carbon atoms, and $a$ is the lattice parameter.  The on-site energy of nickel(cobalt) atoms relative to the $p$-states of carbon is $\Delta_0= \varepsilon_{d_0} -  \varepsilon_p$, whiles $\Delta_{ex}$ represents the exchange energy spin-splitting at the nickel(cobalt) atoms. The parameter  $\eta_{d}$ depends on the overlapping $p_z$-$d$ matrix elements and on the exchange field through the expression,
\begin{equation}
\eta_{d} =\frac{\sqrt{3}a}{2\hbar}\frac{U^{A}_{p_z,z^2}U^{B}_{p_z,z^2}}{\Delta_0^{2}-\Delta_{ex}^{2}},
\end{equation}

\noindent The matrix elements $U^{A/B}_{p_z,z^2}$ are written in terms of the  Slater-Koster parameters $U_{pd\sigma}$ and $U_{pd\pi}$ (see Table I for the estimated graphene/Ni parameters). \cite{shortnotation}  The induced exchange-Rashba $p_z$-$d$ coupling is dictated by the parameter
\begin{equation}
\begin{split}
\lambda_{_{\it XR}}&=\frac{3}{2}\frac{\sqrt{3}b \, \xi_{d} \Delta_{0}\Delta_{1}}{(\Delta^{2}_{1}-\Delta^{2}_{ex})({\Delta_0^{2}-\Delta_{ex}^{2}})}  U^{A}_{p_z,z^2}\widetilde{U}^{B}_{p_z,xz} ,\\
\end{split}
\label{param}
\end{equation}
\noindent 

where $\Delta_1= \varepsilon_{{d_1}} -  \varepsilon_p$, where $\varepsilon_{{d_i}}$, with $i=0,1$ are the relevant onsite energies for Co(Ni), see Appendix A. Notice that the breaking of the inversion symmetry by the Rashba term does not come from a Stark-type effect, as no external electric field is considered here, but from the close proximity to the ferromagnetic atoms instead, due to its relatively large spin-orbit coupling. Indeed, it  is proportional to the intrinsic spin-orbit parameter of nickel(cobalt) $\xi_{d}$, to the effective distance $b$ between  the nickel(cobalt) atom to its nearest-neighbour carbon atom, as well as, to the overlap energies between the $p_{z}$-orbital of graphene atoms of sublattice $B$ with $d_{xz,yz}$ orbitals of nickel(cobalt) atoms, $\widetilde{U}^{B}_{p_z,xz}$; and the overlap of $p_{z}$-orbitals of graphene's atoms of sublattice $A$ and $d_{z^2}$ orbitals of nickel(cobalt) atoms, $U^{A}_{p_z,z^2}$. It also has a nontrivial functional dependence in $\Delta_{ex}$. The last term in Eq.~(\ref{heffpz1}) has two contributions,
\begin{equation}
{  \cal E}_{{ex}}= \varepsilon_{{\it pd}}\otimes s_{y}+\varepsilon^{*}_{{\it D}} \otimes s_{o}.
\end{equation}
\noindent  the first will contribute to a spin-dependent mass term as well as the opening of an energy gap, explicitly $\varepsilon_{{\it pd}}=\Delta_{+}s_o+\Delta_{-}\sigma_{z}$. The second yields a uniform energy shift of the Dirac cones, $\varepsilon^{*}_{D}=\varepsilon_{-}s_o+\varepsilon_{+}\sigma_{z}$, in which
\begin{eqnarray}
2\Delta_{\pm} & = &\left[ \varepsilon_{1}\pm(\varepsilon_{2}+\varepsilon_{3})  \right]\Delta_{ex},\\
2\varepsilon_{_{\pm}} &=&-\varepsilon_{1}\Delta_{0}\pm(\varepsilon_{2}\Delta_{1}+\varepsilon_{3}\Delta_{0}),
\end{eqnarray}

\noindent where we have found that  the dimensionless parameters $\varepsilon_{i}$, ($i=1,2,3$) have the explicit form
\begin{equation}
\begin{split}
\varepsilon_{1}&=\frac{(U^{A}_{p_z,z^2})^{2}}{\Delta_0^{2}-\Delta_{ex}^{2}}, \\
\varepsilon_{2}&=\frac{3b^2} {4( \Delta^{2}_{1}-\Delta^{2}_{ex} ) } \left[  b^2 (\widetilde{U}^{B}_{p_z,xy})^{2} +4(\widetilde{U}^{B}_{p_z,xz})^{2} \right] ,\\
\varepsilon_{3}&=\frac{3(U^{B}_{p_z,z^2})^{2}}{\Delta_0^{2}-\Delta_{ex}^{2}}, \\
\end{split}
\label{heffpzparam}
\end{equation}

\noindent
 in which effective Slater-Koster energies $\widetilde{U}^{B}_{p_z,xy}  $ and $\widetilde{U}^{B}_{p_z,xz}$ depend on the relative orientation of the atomic orbitals (Appendix A).

\subsection{Low Energy Dispersion for  Graphene/Ni}
 
\begin{figure*}
\begin{center}
\includegraphics[width=15.0cm,height=6.0cm]{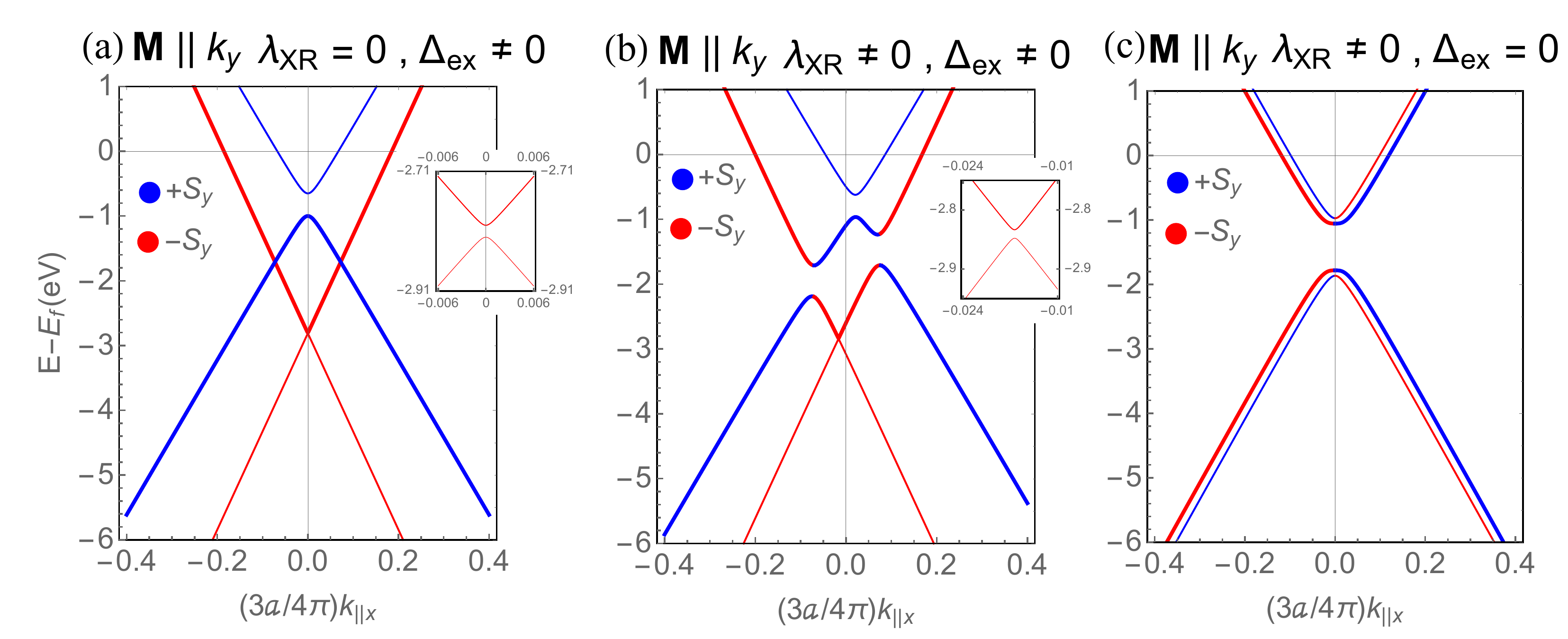}
\end{center}
\caption{\small Band structure in the vicinity of the Dirac point for the graphene-nickel system in the AC stacking configuration for three cases: in (a) the spin-orbit interaction of nickel is neglected and with nonzero exchange interaction, (b)  both the spin-orbit interaction and the exchange (with $\xi_d \sim \Delta_{ex}$) are present, and (c) when we take into account only the spin-orbit interaction with the exchange field zero. In all the cases $k_{||x}$ is the wave vector in the direction $\Gamma-K$, parallel to the graphene's plane, $p_{y}=0$, $p_{x}=\hbar k_{x}-\hbar K_{\xi}$, and  $\xi=+1$. The majority and minority spin bands are indicated with blue and red colors respectively, and the direction of the magnetization of Ni with respect to $k_y$ is indicated. The insets (a) and (b) shows the small gap in the  dispersions for the minority spin bands.}
\label{RpE}
\end{figure*}

The low energy band structure is obtained by diagonalizing the effective Hamiltonian of Eq.~(\ref{heffpz1}) using the tight-binding related parameters contained in Table~\ref{tab1} and ~\ref{tab2}.  Note that for Ni we have $\xi_d \simeq 2\Delta_{ex}$.  However, for illustration purposes we plot in Figure~\ref{RpE} the electronic bands for three different scenarios: {\it (i)} Vanishing spin-orbit coupling ($\xi_{d}=0$) that entails $\lambda_{\it XR}=0$, and  sizable exchange field, $\Delta_{ex}\neq0$,  {\it (ii)}   
 the interplay of both, finite spin-orbit and exchange interactions, and {\it (iii)}
 zero exchange coupling and finite spin-orbit interaction ($\lambda_{\it XR}\neq 0$).\begin{table}
\begin{center}
\caption{Parameters used to calculate the bands structure of Graphene/Ni system. }
\begin{tabular}{ c  c  c }
  \hline
  \hline			
 Parameter 		& Value 		& Ref.  \\
   \hline	 \vspace{-0.25cm}    \\
  $V_{pp\pi}$		& -3.033 eV 	& [\onlinecite{Konschuh}],[\onlinecite{McCann2013}] \\
  $|2\Delta_{ex}|$	& 0.5 eV		& [\onlinecite{Usachov}],[\onlinecite{Varykhalov}],[\onlinecite{Marchenko2015}] \\
  $\xi_{d}$			& 0.1 eV		& [\onlinecite{Barreteau}] \\
  $a$				& 2.46 \AA	& [\onlinecite{McCann2013}] \\
  $d$				& 2.05 \AA	& [\onlinecite{Varykhalov}] \\
 $U_{pd\sigma}$	& -0.651 eV	&[\onlinecite{Inferred}] \\
 $U_{pd\pi}$		& 0.015 eV	&[\onlinecite{Inferred}]  \\
 ${\Delta}_{0}$		& 0.401 eV	&[\onlinecite{Inferred}]\\
 $\Delta_{1}$		& 0.710 eV	&[\onlinecite{Inferred}] \\
    \hline  
    \hline
\label{tab1}
\end{tabular}
\end{center}
\end{table}

In the first scenario ($\lambda_{\it XR}=0$) the corresponding dispersion laws for the  majority/minority bands follows,
\begin{equation}
\begin{split}
E_{\pm\uparrow} ( p ) &= {\cal E}_{D} +\Delta_{+}\pm \sqrt{(\varepsilon_{+} + \Delta_{-})^2 + (v^{*}_F - v^{*}_{d})^2 p^2},\\
E_{\pm\downarrow} ( p )  &={\cal E}_{D} -\Delta_{+}\pm \sqrt{(\varepsilon_{+} - \Delta_{-})^2 + (v^{*}_F + v^{*}_{d})^2 p^2}.\\
\end{split}
\label{eigenvalues1}
\end{equation}

\noindent where ${\cal E}_{D}=\varepsilon_{-}$, and  the $\pm $ sign in $E_{\pm,\uparrow\downarrow}$ corresponds to the electron/hole bands, respectively. In Fig.~\ref{RpE}(a) we have plotted these bands and observe that the behavior of  the dispersions for minority spin-bands (in red) preserves the linear in momentum behavior at the vicinity of the Dirac point, with a small gap of $2|\varepsilon_{+}-\Delta_{-}|=20$\,meV. These bands are symmetric with respect to the Dirac point, shifting the minority spin-bands towards lower energies by  $E_{\pm \downarrow} ( 0 )= {\cal{E}}_{D} -\Delta_{+}$=-2.8\,eV, according to the experimental value reported of $\approx 2.82$eV for graphene/Ni in the AC stacking.\cite{Varykhalov, Usachov} In contrast, a gap of $350$\,meV is opened for the majority spin-bands (in blue). This gap is governed by $ | {E}_{+\uparrow}   ( 0 ) - {E}_{-\uparrow}  ( 0 ) |=2|\Delta_{-}+\varepsilon_{+}|$.   Also a rather large exchange spin-splitting separation between the majority hole band and the minority electron band is predicted, corresponding to an exchange energy of $ | {E}_{-\uparrow}   ( 0 ) - {E}_{+\downarrow}  ( 0 ) |=2|\Delta_{+}-\varepsilon_{+}|=1.8$\,eV.

Notice  that we can assign a spin dependent velocity of the spin-bands, as $v_{F\uparrow(\downarrow)}=v^{*}_F \pm v^{*}_{d}$, that makes the velocities slightly different for the majority and the minority spin bands (see Table~\ref{tab2}). Moreover, as can be shown from Eq.(\ref{eigenvalues1}) the term $\varepsilon_{+} \pm \Delta_{-}$ introduces  in general a spin-dependent effective mass term  that determines the opening of a gap between the majority/minority electron and hole bands.  Since in this case we have that $\varepsilon_{+} \simeq \Delta_{-}$, (see Table~\ref{tab2}) then or the minority spin bands the gap vanishes, red lines in Fig. \ref{RpE}(a).

For the most general case ({\it ii}) in which the spin-orbit interaction is included and the interplay of the exchange coupling is allowed, there is no a simple analytical solution for eigenvalues of  Eq.~(\ref{heffpz1}).  In Figure~\ref{RpE}(b) we show the numerical solution for the bandstructure for such case. The electron and hole bands become asymmetric with respect to the Dirac point, appearing large anticrossing gaps at $k_{||x}\approx \pm 0.128$\,\AA $^{-1}$  of about $450$\,meV (Figure~\ref{RpE}(b)). We can see also that the spin polarization along the $y$-direction $\langle   S_y \rangle _{\pm}$ changes around these values of $k_{||x}$ for the two adjacent bands that form the anticrossing gaps (thick lines). In Table~\ref{tab2} we show the values of the parameters used for this case.
\begin{table}
\vspace{-0.5cm}
\begin{center}
\caption{Estimated parameters.}
\begin{tabular}{ c  c  c}
  \hline
  \hline      
 Parameter            			& $\xi_d \sim \Delta_{ex}$(Ni)   	& $\xi_d \ll \Delta_{ex}$(Co)   \\
 \hline	   \\
  $\varepsilon_{-}$   			& -1.8175 eV   				& -1.470 eV   \\
  $\varepsilon_{+}$   			& 0.0925 eV   				& 0.030 eV \\
  $\Delta_{+}$        			& 0.9925 eV  				& 1.330 eV   \\
  $\Delta_{-}$       	 		& 0.0825 eV  				& 0.0298 eV    \\
  $\lambda_{\it XR}$      		& 0.242 eV  					& 0.265 eV  \\
  $v^{*}_{F}$             			& 1.21$\times10^6$ m/s   		& 1.026$\times10^6$ m/s\\
  $v^{*}_{d}$         			& 143196 m/s   				& 42749 m/s \\
  $U^{A}_{p_z,z^2}$   		& -0.651 eV  				& -0.363 eV \\
  $U^{B}_{p_z,z^2}$   		& -0.268 eV 				& 0.035eV  \\
  $\widetilde{U}^{B}_{p_z,xz}$	& -0.767 eV   				& -0.538 eV\\
  $\widetilde{U}^{B}_{p_z,xy}$	& -0.951 eV  				& -0.985  eV\\
  $b$ 					& 0.569  					& 0.558 \\
  \hline 
  \hline
  \label{tab2}
\end{tabular}
\end{center}
\vspace{-1cm}
\end{table}
 The spin-orbit induced anticrossing gap for the electron/hole ($\nu=e,h$) bands at the extremum with momentum $k_{m}$ is defined as $\Delta_{so}^{\scriptscriptstyle (\nu)}(k_{m}) = |E_{\scriptscriptstyle {\nu,+}} (k_{m})-E_{\scriptscriptstyle {\nu,-}} (k_{m}) |$. For the case in Figure~\ref{RpE}(b)  we have $k_m=0.128$\,\AA $^{-1}$. Clearly there is an asymmetric spin-splitting as $\Delta_{so}^{\scriptscriptstyle (\nu)}(k_{m})\neq\Delta_{so}^{\scriptscriptstyle (\nu)}(-k_{m})$. The Dirac point for the hole-minority band is shifted to a negative value by $k_o =-0.03$\,\AA $^{-1}$. This asymmetry of the band structure with respect to the Dirac point, is characteristic of the joint presence of induced Rashba and exchange couplings in graphene, as reported by Rybkin {\it et al.}\cite{Rybkin} for graphene/Au/Co. In such work the authors observed spin-orbit splittings of the order of $200\pm40$\,meV at the $\it{K}$ point, and about $40\pm40$\,meV in the $\it{K'}$ point, measurements with a reasonable agreement with their DFT calculations. Here we illustrate that {\bf a} similar effect can likewise occur in graphene/Ni.

We consider now  the third scenario, vanishing exchange field ($\Delta_{ex}=0$) with finite spin-orbit interaction ($\lambda_{\it XR}\neq 0$) which leads also to analytical expressions for the electronic bands, $E_{c/v,s}  ( p ) =  {\cal E}_{D} \pm {\cal E}_{s} ( p )$, with
\begin{equation}
{\cal E}_{s} ( p ) =  \sqrt{\varepsilon^{2}_{+} + \left(\lambda_{\it XR} +s\sqrt{ \lambda^{2}_{\it XR}+( v^{*}_F p)^2  } \right)^2},
\end{equation}

\noindent where $\{c,v\}$ denotes the electron/hole branches, respectively,  and the label $s=\pm$ refers to the spin helicity. In Figure~\ref{RpE}(c) we plot the dispersion bands. Clearly the bands recover the symmetry with respect to the Dirac point.  The sign of spin polarization $S_y$ changes when $k_{\parallel x}\rightarrow-k_{\parallel x} $,  in accordance with standard notion of Rashba spin-orbit coupling. The Rashba spin-splitting energy at $p=0$ is thus given by $ \Delta_R \equiv | {\cal E}_{+}   ( 0 ) - {\cal E}_{-}  ( 0 ) |$, with 
\begin{equation}
\Delta_R  = \sqrt{\varepsilon^{2}_{+} + 4 \lambda_{\it XR}^2 }-\varepsilon_{+},
\end{equation}

\noindent  leading to  $\Delta_R=82.8$ meV (see Table~\ref{tab3}).  Due the $p_z$-$d$ hybridization between cobalt atoms with graphene, a rather huge gap is opened between the bands with spin-chirality $s=-$, and is given by $| E_{c,-}  ( 0 ) - E_{v,-}  ( 0 ) | =2|\varepsilon_{+} |$, corresponding to an energy of about  $0.73$\,eV. 
 
\begin{figure*}
\begin{center}
\includegraphics[width=15.0cm,height=6.0cm]{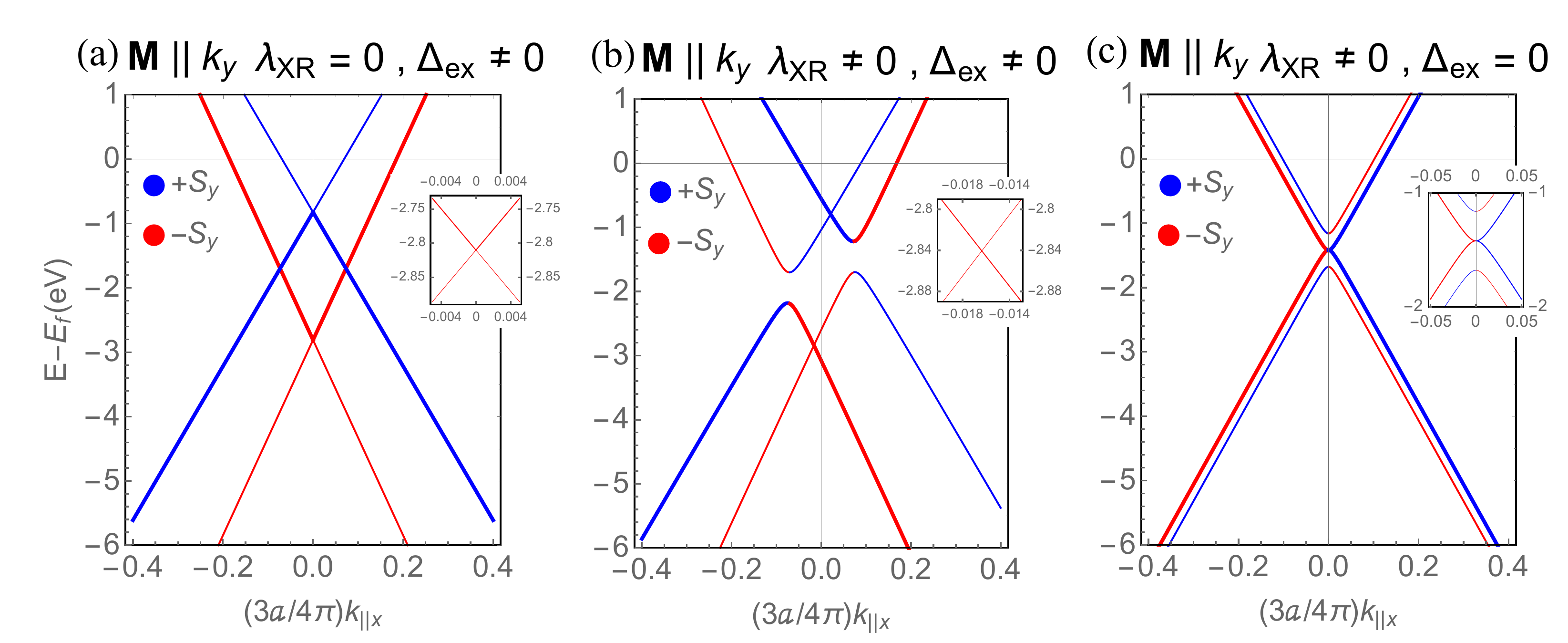}
\end{center}
\caption{\small Band structure in the vicinity of the $K$ point as a function of  $k_{||x}$  with $k_{y}=0$ for the graphene/Ni neglecting the small parameters $\Delta_{-}$ and $\varepsilon_{+}$.   We show here the same three cases of Fig.~\ref{RpE}: (a) finite exchange coupling in the absence of the spin-orbit interaction of the Ni-atoms, (b) when both, the spin orbit interaction and the exchange are present,  and (c) with spin-orbit interaction present whiles and the exchange field is set to zero.  The insets show the absence of gaps between the bands in (b) and the Rashba splitting in (c). 
 }
\label{RpEap}
\end{figure*}
Figure~\ref{RpEap} shows the same three scenarios studied in Figure~\ref{RpE} but  neglecting the parameters $\Delta_{-}$ and $\varepsilon_{+}$, which are small in magnitude in comparison to $\Delta_{+}$ and $\varepsilon_{-}$, respectively  (see Table~\ref{tab2}). The main difference with respect to the results shown in Figure~\ref{RpE}  is the closing of the gaps between the majority and the minority spin bands for the cases where it was assumed either $ \lambda_{\it XR }\neq0$ or $ \lambda_{\it XR }=0$, with $\Delta_{ex}\neq0$ (Figure~\ref{RpEap}(a),(b)). Since $\varepsilon_{+}$ is taken equal to zero then $| E_{c,-}  ( 0 ) - E_{v,-}  ( 0 ) | = 0$ leading to a gapless bandstructure, Figure~\ref{RpEap}(c).
\begin{table}
\begin{center}
\caption{Hamiltonian parameters for $\Delta=0$ (case $\xi_d \sim \Delta$).}
\begin{tabular}{ c  c }
  \hline \hline     
 Parameter            & Value      \\
  \hline  
  $\varepsilon_{-}$   	& -1.428 eV     \\
  $\varepsilon_{+}$   	& 0.373 eV    \\
  $\Delta_{+}$        	& 0   \\
  $\Delta_{-}$        	& 0    \\
  $\lambda_{\it XR}$	& 0.136 eV  \\
  $v^{*}_F$             	& 1.126$\times10^6$ m/s   \\
  $v^{*}_{d}$         	& 0   \\
  \hline  \hline
  \label{tab3}
\end{tabular}
\end{center}
\end{table}
For all the cases presented in Figures~\ref{RpE} and \ref{RpEap}, we find that the Fermi velocities of the bands near the Dirac point ranges between $0.9\times 10^6$\,m/s and $1.1\times 10^6$\,m/s (see Tables~\ref{tab2} and \ref{tab3}), which are very close to the experimental Fermi velocities for the intact Dirac cones reported for Graphene/Ni ($\approx 0.8\times 10^6$m/s \cite{Varykhalov, Usachov}). 

\begin{figure}
\begin{center}
\includegraphics[width=9.0cm,height=4.5cm]{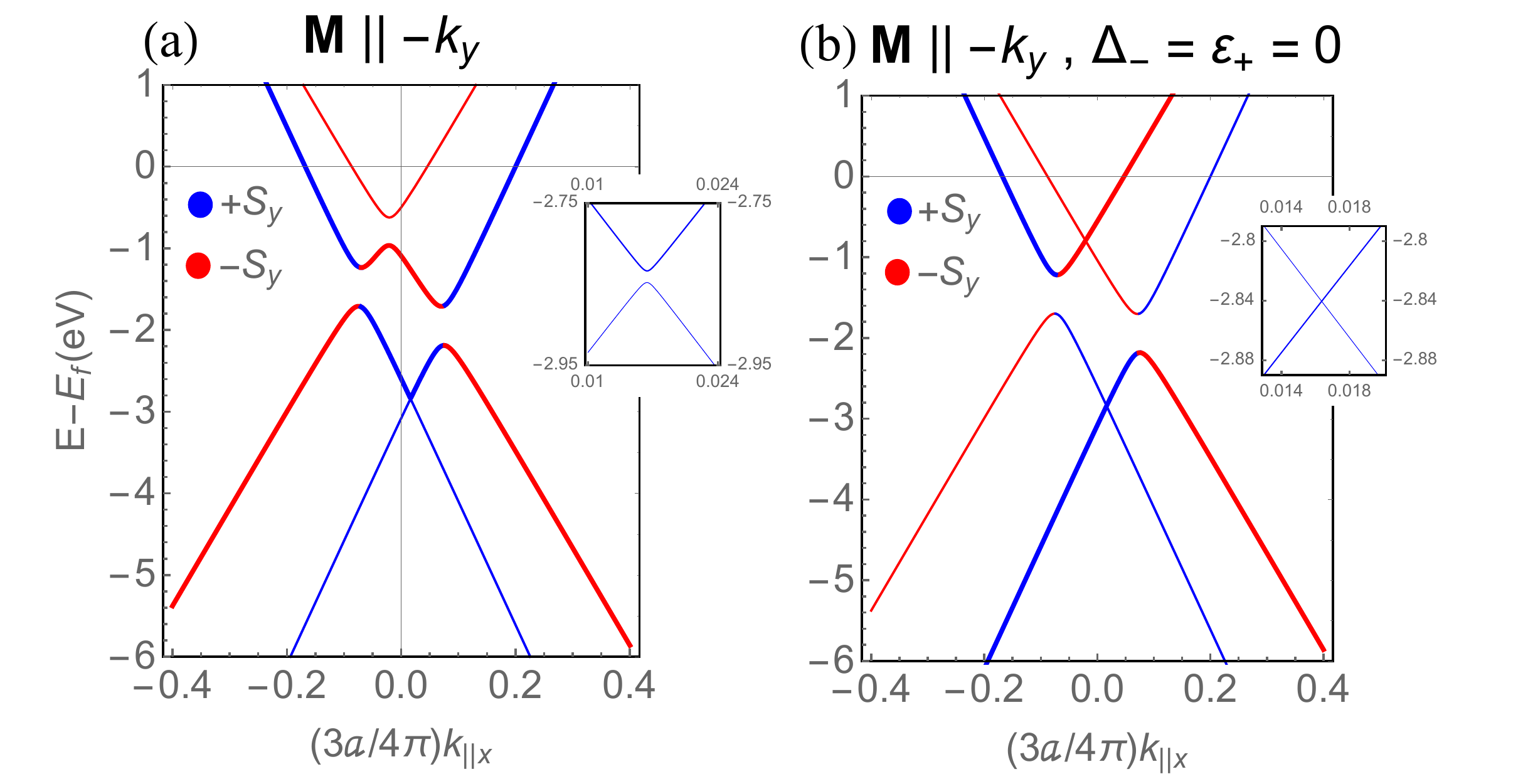}
\end{center}
\caption{\small (Band structure in the vicinity of the Dirac point for the graphene-cobalt with inverted magnetization respect to Figure~\ref{RpEap} for: (a) when we take the parameters $\Delta_{-}$ and $\varepsilon_{+}$, (b) when we neglect the parameters $\Delta_{-}$ and $\varepsilon_{+}$. In both the cases $k_{||x}$ is the wave vector in the direction $\Gamma-K$, parallel to the graphene's plane, $p_{y}=0$, $p_{x}=\hbar k_{x}-\hbar K_{\xi}$, and  $\xi=+1$. The majority and minority spin bands are indicated with blue and red colors respectively, and the direction of the magnetization of Co with respect to $k_y$ is indicated. The insets show the gaps between the bands in (a) and the absence of gaps in (b).}
\label{RpEinvmag}
\end{figure}

The behavior of the band structure upon the inversion of the in-plane magnetization vector of the Ni layer (parallel to -$k_y$) is also studied and it is shown in Figure~\ref{RpEinvmag}. We plot the most general scenario in which we have the joint action of the spin-orbit interaction  and the exchange field (Fig.~\ref{RpEinvmag}(a)), as well as the case in which the parameters $\Delta_{-}=\varepsilon_{+}=0$, Fig.~\ref{RpEinvmag}(b). As expected,  if the magnetization in nickel is inverted, the spin polarization of  the graphene bands is likewise inverted. The asymmetries of the bands with respect to the Dirac point become also the mirror images with respect to the magnetization parallel to $+k_y$. 

\subsection{Low Energy Dispersion for  Graphene/Co} 

\begin{figure}
\begin{center}
\vspace{-1.5cm}
\includegraphics[width=9.0cm,height=7.0cm]{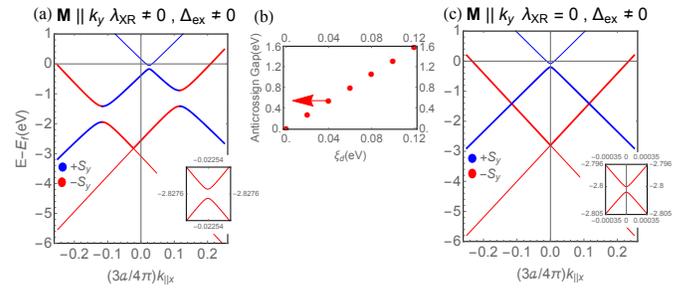}
\end{center}
\vspace{-2cm}
\caption{\small (a): Band  structure in the vicinity of the Dirac point with $k_{y}=0$ for the graphene-cobalt system in the AC stacking configuration, for the case in which the spin-orbit interaction of cobalt is included. As before, $k_{||}$ is the wave vector in the direction $\Gamma-K$, parallel to the graphene's plane. The majority and minority spin bands are indicated with blue and red colors respectively. (b) Plot of the anti-crossing gap in function of the tight-binding spin-orbit parameter $\xi_{d}$. (c) same as (a) but neglecting the spin-orbit coupling.}
\label{eigenvalues2}
\end{figure}
Interestingly, our Slater-Koster tight-binding model offers an explanation for the appearance of intact and gapless Dirac cones formed by the minority spin bands in graphene on cobalt, as observed recently through spin- and angle-resolved photoemission measurements. \cite{Varykhalov, Marchenko2015, Usachov} We notice that the preservation of the linearity of the spin-bands together with a gapless (or nearly gapless) regime occurs whenever the atomic spin-orbit coupling is much smaller than the exchange field ($\xi_d \ll \Delta_{ex}$) as for the case of graphene/Co (Table~\ref{tab4}). 
In Figure~\ref{eigenvalues2}(a) we plot the resulting band structure for graphene/Co. 
We observe that the minority spin bands  show a linear dispersion at the vicinity of the $K$ point with a rather small gap $\lesssim 1$\,meV, in contrast with the case of graphene/Ni where $\xi_d \sim \Delta_{ex}$ and a gap of about 20 meV is obtained. The  smallness of such gap in graphene/Co turns to be much smaller than the precision of the SARPES experimental reports on gapless intact Dirac cones. \cite{Varykhalov, Marchenko2015, Usachov}  We may presume that such tiny gap predicted here was actually present in these experiments, but very likely passed unnoticed due their lack of resolution for such small range of energies and the broadening of the bands. We also observe that the asymmetry generated by the spin-orbit interaction is not as pronounced as in the $\xi_d \sim \Delta_{ex}$ case. Here the horizontal uniform asymmetric momentum shifting of the bands is also of the order  $(3a/4\pi)k_{\parallel x}=0.02$.  Notice that the model also predicts the opening of a rather large anticrossing gaps between the majority and minority spin bands ($\sim$ 0.510\,eV) , here occuring at approximately  $\pm 0.19$\AA $^{-1}$.
Such gaps were not reported in the experimental bands for Graphene/Co(0001)\cite{Varykhalov, Marchenko2015, Usachov}, perhaps due the fact that these bands were not clearly resolved in the experiment below the Dirac point due to the strong hybridization of the $\pi$ bands of graphene with the $3d$ bands of cobalt.  We can trace-back the origin of this effect to the induced magneto-spin-orbit Rashba type term in the low energy effective Hamiltonian (Eq.~(1)) and governed by the strength of $\lambda_{_{XR}}$.  In Figure~\ref{eigenvalues2}(b) we depict the linear relationship between these anticrossing gaps and the intrinsic spin-orbit parameter $\xi_{d}$ of cobalt. The opening of such gaps arises clearly as long as $\xi_d \sim \Delta_{ex}$, exhibiting  closing gaps for the physical condition for which $\xi_d \ll \Delta_{ex}$.  The effect of neglecting the $\lambda_{_{XR}}$ contribution  leading to vanishing anticrossing gaps far away from the Dirac point  is shown in Figure~\ref{eigenvalues2}(c).

\begin{table}
\begin{center}
\caption{Parameters used to calculate the bands structure of Graphene/Co system} 
\begin{tabular}{ c  c  c }
  \hline\hline			
 Parameter 		& Value 		& Ref. \\
  \hline	
  $V_{pp\pi}$		& -3.033 eV 	&[\onlinecite{Konschuh}],[\onlinecite{McCann2013}] \\
  $|2\Delta_{ex}|$	& 1.6 eV		&[\onlinecite{Usachov}],[\onlinecite{Varykhalov}],[\onlinecite{Marchenko2015}]\\
  $\xi_{d}$			& 0.04 eV		&[\onlinecite{Barreteau}]\\
  $a$				& 2.46 \AA	&[\onlinecite{McCann2013}] \\
  $d$				& 2.11\AA		&[\onlinecite{Varykhalov}]\\
  $U_{pd\sigma}$	& -0.126 eV	&[\onlinecite{Inferred}] \\
  $U_{pd\pi}$		& 0.286 eV	&[\onlinecite{Inferred}] \\
  ${\Delta}_{0}$		& -0.794 eV	&[\onlinecite{Inferred}]\\
  $\Delta_{1}$		& 1.246 eV	&[\onlinecite{Inferred}] \\
  \hline \hline 
\label{tab4}
\end{tabular}
\end{center}
\end{table}

\begin{figure}
\begin{center}
\includegraphics[width=9.0cm,height=7.0cm]{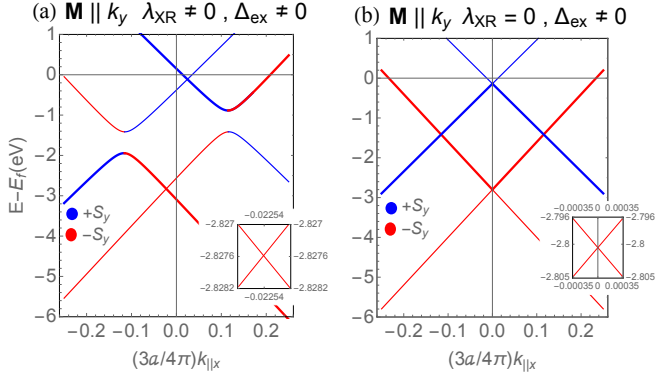}
\vspace{-2cm}
\end{center}
\caption{\small Bands structure in the vicinity of the Dirac point for the graphene-cobalt system in the AC stacking configuration: (a)  neglecting the parameters $\varepsilon_{+}$ and $\Delta_{-}$, and (b) {\bf same as (a), but} for the case in which the spin orbit interaction of cobalt is not included. In both plots $k_{||}$ is the wave vector in the direction $\Gamma-K$, parallel to the graphene's plane. In these plots we took $p_{y}=0$, $p_{x}=\hbar k_{x}-\hbar K_{\xi}$, and  $\xi=+1$. The majority and minority spin bands are indicated with blue and red colors respectively.}
\label{eigenvalues3}
\end{figure}

In order to further illustrate the strong influence in the anticrossing effect of the cobalt spin-orbit coupling $\xi_d$ as well as the sensitivity of the parameters on the band structure, we plot in Figure~\ref{eigenvalues3} the bands for the system graphene/cobalt  in the limit $\varepsilon= 0$ and $\Delta_{-}= 0$,  for two cases: Figure~\ref{eigenvalues3}(a) with a nonzero spin-orbit interaction term (thereby $\lambda_{_{XR}}\neq0$) , and in Figure~\ref{eigenvalues3}(b) when the spin-orbit coupling is totally absent (Figure~\ref{eigenvalues3}(b)) but maintaining the rest of the parameters as in the case $\xi_d \ll \Delta_{ex}$ (Table \ref{tab4}).  The closing of the gap at $(4\pi/3a)k_{\parallel,x}\simeq 0.1$ by making $\xi_d=0$  is evident (Figure~\ref{eigenvalues3}(b)). Interestingly  under this conditions a gapless intact Dirac cone is formed by the minority spin bands with its apex shifted by  $E_{\pm \downarrow} ( 0 )= {\cal E}_D-\Delta_{+}=-2.8$\,eV.  The effective masses associated to the spin- dispersions laws of Eq.(\ref{eigenvalues1}) are dictated by  

\begin{equation}
\frac{1}{m_{\pm\uparrow,\downarrow}^{*}}=\pm\frac{(v_{F\uparrow,\downarrow}^{*})^2}{\varepsilon_{+} \pm \Delta_{-}},
\end{equation}

\noindent where explicitly in  terms of the Slater-Koster parameters we have
\begin{equation}
\begin{split}
\varepsilon_{+} \pm \Delta_{-}& = \frac{3b^2}{2} \frac{ (\widetilde{U}^{B}_{p_z,xz})^{2} +\frac{1}{4}b^{2}(\widetilde{U}^{B}_{p_z,xy})^{2} }  {(\Delta_{1}\pm\Delta_{ex})} \\ &+\frac{3}{2}\frac{(U^{B}_{p_z,z^2})^{2} - \frac{1}{3}(U^{A}_{p_z,z^2})^{2}}{2({\Delta}_{0}\pm\Delta_{ex})},\\
\label{effectivemass}
\end{split}
\end{equation}

\noindent which gives insight to the type of Co-graphene hibridizations that determines the breaking of the Dirac cones and the appearing of parabolic bands, as well as the criteria for the preservation of massless Dirac Fermions for the spin-minority bands. For instance, vanishing effective mass for the spin-minority bands  entails $(\varepsilon_{+}- \Delta_{-})\rightarrow 0$, physical  situation occurring in Fig.\,(\ref{eigenvalues3}). Indeed, we can corroborate from Eq.\,(\ref{effectivemass}) and Table~\ref{tab2}    that the first term to the right basically cancels the second term, producing massless intact Dirac cones.  It is worth noticing that due the $p$-$d$ orbital overlapping in combination with the exchange field, the Fermi velocities become renormalized such that the velocity  $ v^{*}_{\downarrow}$ for the spin minority linear bands becomes reduced respect to the free-standing graphene $v_F$.

\section{Conclusions}
 
We investigated theoretically the proximity effects induced in a monolayer of graphene over a monolayer of  a ferromagnetic metal Ni and Co within the Slater-Koster approximation in a multi-orbital tight-binding approach. We construct a minimal model that includes the spin-orbit coupling and the exchange fields of the ferromagnetic atoms,  besides the hybridizations of the $3d$-orbitals of Ni(Co) atoms with those of graphene's $p_z$-orbitals. From this, we derive a low energy effective Hamiltonian that dictates the dispersion laws for graphene/Ni(Co) which is consistent with recent DFT calculations and with photoemission experiments with spin-resolution.

The low energy Hamiltonian contains three dominant terms: an effective Dirac term with a perturbed velocity shifted by an exchange field dependent velocity.  The second term has formally the structure of a Rashba type  spin-orbit interaction, transferred to graphene from the intrinsic spin-orbit of nickel(cobalt) and from the overlaps between the A(B) $p_z$ orbitals of graphene with the $d_{z^2}$($d_{xz}$ and $d_{yz}$) of the ferromagnetic atoms. Finally the third term has the form of a constant matrix that mostly depends on the exchange fields.
 
 As a consequence of the nature of these terms in the effective Hamiltonian, the first which is linear in momentum, produces spin-asymmetric bands with respect to $k_{||x}=0$ with reduced Fermi velocities. Then the joint action of the Rahba-type and exchange terms induces a magneto-spin-orbit effect on the spin bands, leading to large asymmetries between the minority(majority) spin bands and the appearance of gaps a the Dirac point and to anticrossing-gaps away from it. 
 
We illustrated two regimes. In the first, the spin orbit coupling is similar to the exchange coupling ( $\xi_d \sim \Delta_{ex}$). This situation physically corresponds to the system of graphene/Ni in the AC stacking configuration. We showed that the band structure for this system presents the combination of a strong exchange and Rashba coupling, known as the Rashba+Exchange effect.\cite{Rybkin} This description is given in terms of the overlaps between orbitals, the exchange coupling and the atomic spin orbit parameter of nickel.  The second regime corresponds to the case in which the spin orbit coupling is much smaller than the exchange coupling ($\xi_d \ll \Delta_{ex}$). This scenario occurs in the system of graphene/Co in the AC stacking configuration. For this system, we were able to give an insight on the experimentally observed intact Dirac cones.\cite{Varykhalov, Marchenko2015, Usachov} In this sense, we noticed that this effect is produced because the spin dependent effective mass term, which depends on the graphene-cobalt hybridizations, goes to zero for the minority spin bands. In summary, we have shown that proximity effects emerging from depositing graphene on a ferromagnetic metal substrate such as nickel and cobalt can change significantly the dynamics of its Dirac electronic states, inducing very large exchange fields and giant spin-orbit coupling to the graphene layer comprising a magneto-spin-orbit effect and intact Dirac bands which may be of great interest for spintronics applications.  
 
 \acknowledgements
M. P.  acknowledges the use of the facilities of the Center of  Nanosciences and Nanotechnology (CNyN) of the National Autonomous University of Mexico, as well as to the National Council for Science and Technology of Mexico (CONACyT) and to the Mexican ministry of energy (SER) for providing a post-doctoral fellowship to carry out this research.

\appendix

\section{Tight-binding approach}\label{TB}
In the absence of spin-orbit and exchange interactions the Hamiltonian for graphene/Ni(Co)  interface in real space reads
\begin{equation}
\label{H}
\hat{H}_{_{TB}}=-\frac{\hbar^2}{2m}\nabla^2+\sum_{i}V_{_{C}}(\bm{r}-\bm{R}^{C}_{i})+\sum_{i'}V_{_{Ni(Co)}}(\bm{r}-\bm{R}^{Ni(Co)}_{i'}),
\end{equation}

\noindent where the first term corresponds to the kinetic energy of the electrons and the second/third terms to the atomic potential $V_{_{C/Ni(Co)}}$ of carbon/nickel(cobalt) ions, which in the two center approximation, only involves the atoms $i/i'$ at the vector positions $\bm{R}^{C}_{i}/\bm{R}^{Ni(Co)}_{i'}$ and its first neighbors, respectively. We are interested in the matrix elements of $\hat{H}_{_{TB}}$ in the subspace spanned by the  $2p_z$ orbitals for the two distinct $A$ and $B$ sites of graphene  with states $|p^{A/B}_{z,s}\rangle$ and in the subspace spanned by the $3d$-orbitals of Ni(Co) (only at sites $A$), that is,  the states $|\phi_{\mu}\rangle\otimes|s\rangle$, where $\mu=\{1,2,3,4,5\}$ denotes the five $3d$-orbitals of cobalt, and $s=\{\uparrow,\downarrow\}$ labels the electron spin. 
The Wannier functions centered on the different atomic sites are assumed to be orthogonal. Hence the Slater-Koster Hamiltonian matrix elements are $\langle\varphi_{\mu,s}|\hat{H}_{_{TB}}|\varphi_{\mu',s'} \rangle$, with $\{|\varphi_{\mu,s}\rangle\}=\{|p^{A}_{z,s}\rangle, |p^{B}_{z,s}\rangle, |\phi_{\mu,s}\rangle\}$.  This generates a $14\times14$ overlapping matrix for the graphene/Ni(Co) interface, 

\begin{equation}
 H_o =
\left(\begin{array}{ c  c }
  H_\pi					& U \\ 
  U^{\dagger}  	 		&{\cal{E}}_{_{Ni(Co)}} \\   		   
\end{array}\right)\otimes \{\uparrow,\downarrow\} ,
\label{h}
\end{equation}
where $H_{\pi}$ is the bare graphene Hamiltonian matrix in the basis $\{p^{A}_{z},p^{B}_{z}\} \otimes \{\uparrow,\downarrow\}$, having the form
\begin{equation}
H_{\pi} =
\left(\begin{array}{ c c }
  0					& V_{pp\pi} \\ 
  V_{pp\pi}  	 		& 0 \\  		   
\end{array}\right),
\label{hgr}
\end{equation}
\noindent and the on-site energy of the $2p$ orbitals of graphene, $\varepsilon_{p}=\langle p^{A/B}_{z,s}|\hat{H}_{_{TB}}|p^{A/B}_{z,s}\rangle=0$,  has being chosen as the reference energy. Here $V_{pp\pi}$ is the Slater-Koster parameter for the overlaps between $p_z$ carbon orbitals when they form pure $\pi$-bondings. \cite{SlaterKoster} 

On the other hand,  for the Ni(Co)-atoms there is a natural breaking of the degeneracies of the $3d$-orbital states due to the crystal field. Hence, in the absence of magnetic fields, such that time reversal symmetry is satisfied,  the five  $3d$-orbitals of Ni(Co) split into three spin degenerate groups: two doublets $E_1$ (with $3d_{xz}$ and $3d_{yz}$  orbitals) and $E_2$ (with $d_{xy}$ and $3d_{x^2-y^2}$ orbitals ) and a singlet $A_1$ (consisting only of $3d_{z^2-r^2}$ orbitals). \cite{Kirczenow} In the basis $\{|\phi_{\nu}\rangle\}=\{d_{xz},d_{yz},d_{3z^2-r^2},d_{xy},d_{x^2-y^2}\}$, the on-site matrix elements for the Ni(Co)-atoms with respect to the on-site energy $ \varepsilon_p$ of the $p$ orbitals of graphene is given by

\begin{equation}
{\cal{E}}_{_{Ni(Co)}} =
\left(\begin{array}{ c c c c c  }
  \varepsilon_{d_1} -  \varepsilon_p					& 0    & 0  &   0 & 0 \\ 
  0  	 		&                  \varepsilon_{d_1} -  \varepsilon_p	    & 0    & 0  &   0 \\  
   0  	 		&                 0    &  \varepsilon_{d_0} -  \varepsilon_p	   & 0  &   0 \\  
    0  	 		&                 0    & 0    &  \varepsilon_{d_2}-  \varepsilon_p	  &   0 \\  
     0  	 		&                 0    & 0    & 0  &    \varepsilon_{d_2} -  \varepsilon_p	 \\  		   
\end{array}\right),
\label{hCo}
\end{equation}

\noindent where the label $d_{|m|}$ correspond to the $m=0,\pm1,\pm2$ magnetic orbital states of the on-site energies $\varepsilon_{d_{|m|}}$. Here we are assuming that the energetically favorable configuration is that with Ni(Co)-atoms  lying directly underneath the graphene sublattice $A$. Therefore, it is expected that the $3d_{3z^2-r^2}$  orbitals of the $A_1$ group (referred now on as $d_{z^2}$ orbitals) will hybridizes strongly with $p_{z}$ orbitals of the sublattice $A$ carbon atoms. By the same token, the Ni(Co)-atoms of the $E_{1,2}$ group symmetry containing the orbitals $d_{\{xz,yz,xy,x^2-y^2\}}$ will not hybridize with carbon atoms of the $A$ sublattice, but only with the sublattice $B$. Given the marked differences between the hybridization of these two groups of orbitals, here we shall assume that the onsite energies are renormalized by the proximity with graphene such that $\varepsilon_{d_2}\simeq\varepsilon_{d_1}$ \cite{Varykhalov}.

\begin{figure}
\begin{center}
\includegraphics[width=7.5cm,height=5.0cm]{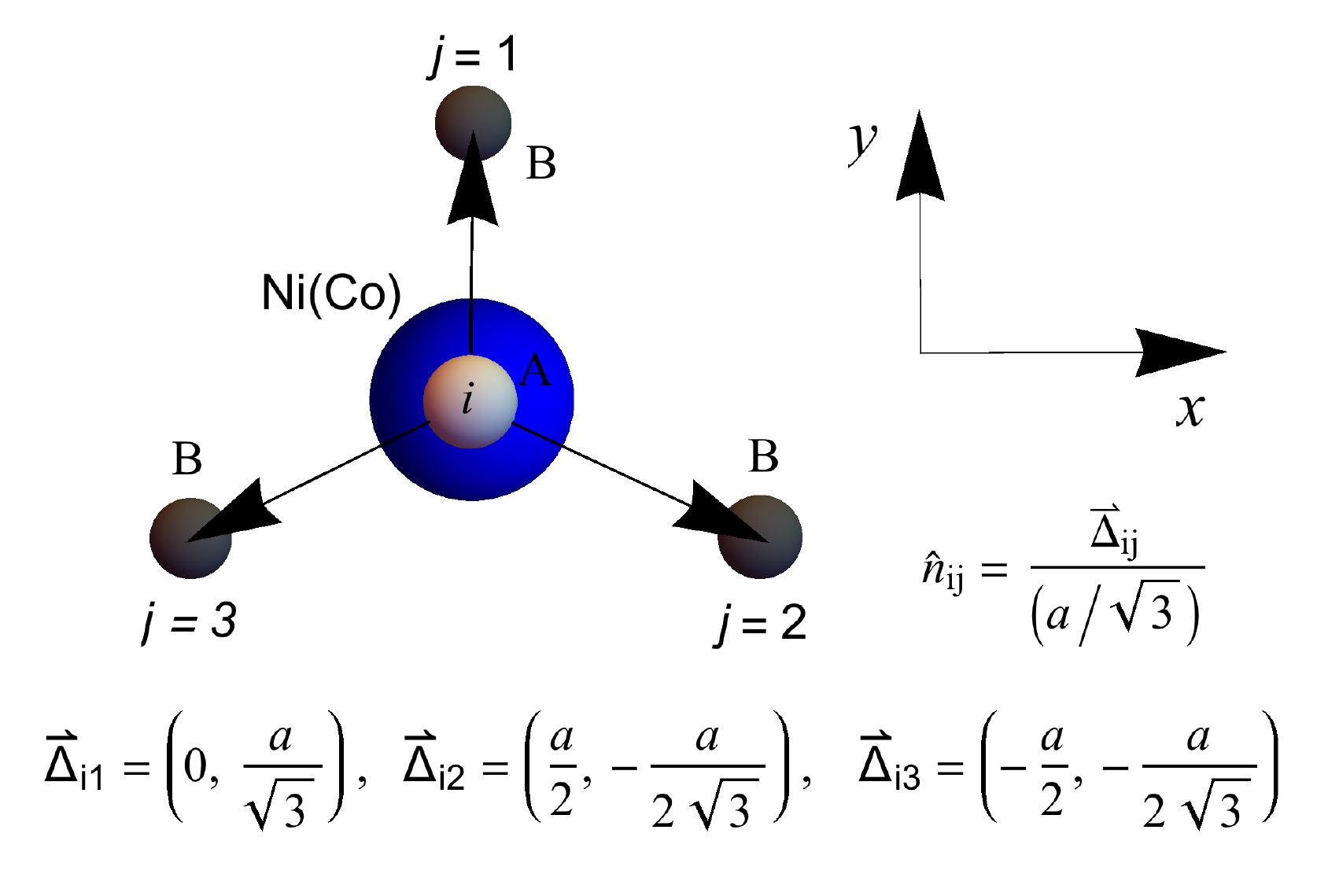}
\end{center}
\caption{\small Scheme that shows the cosine directors to first neighbors, from atoms at $A$ sublattice to atoms to $ B$ sublattice in the graphene plane, $n_{ij}$. The cosine directors from the $i$-th carbon atom at $A$ to the nearby nickel(cobalt) atom are, $\nu_{ix}=0$, $\nu_{iy}=0$ and $\nu_{iz}=-1$; whiles the cosine directors from nearest neighbor $j=1,2,3$ carbon atoms at  $B$  to the nickel(cobalt) atom are defined by $\mu_{jx}=-b n_{ijx}$, $\mu_{jy}=-b n_{ijy}$ and $\mu_{jz}=d/\zeta$, where  $b=(a/\sqrt{3})/\zeta$, with $\zeta =\sqrt{(a/\sqrt{3})^2+d^2}$, (not shown). Here the graphene lattice parameter $a=2.46$\AA \,, and the graphene-Ni(Co) distance is $d$.}
\label{reciprocalvec}
\end{figure}

The graphene-nickel(cobalt) overlaping matrix $U$, reads
\begin{equation}
U(\hat\tau_{j}) =\left(\begin{array}{ c c c c c }
   U^{A}_{p_z,xz}      	&  U^{A}_{p_z,yz} 		&   U^{A}_{p_z,z^2}            &  U^{A}_{p_z,xy}    		&  U^{A}_{p_z,x^2}	\\
   U^{B}_{p_z,xz}      	&  U^{B}_{p_z,yz} 		&   U^{B}_{p_z,z^2}            &  U^{B}_{p_z,xy}    		&  U^{B}_{p_z,x^2}	\\	   
\end{array}\right),
\label{u}
\end{equation}
\noindent where the matrix elements $U^{A/B}_{p_z,\nu} \equiv \langle p^{A/B}_{z,s}|\hat{H}_{_{TB}}|\phi^{Ni(Co)}_{\nu,s}\rangle $ depend explicitly on the Slater-Koster parameters $V_{pd\sigma}$ and $V_{pd\pi}$  corresponding to overlaps between carbon $p_z$ orbitals and nickel(cobalt) $d$ orbitals, and the cosine directors joining these atoms (see Fig. (\ref{reciprocalvec})). 

\begin{widetext}
\begin{equation}
\begin{split}
U^{A/B}_{p_z,z^2}	&= \langle p^{A/B}_{z,s}|\hat{H}_{_{TB}}|d_{3z^2-r^2,s}\rangle =  \sqrt{3} m_{z}(m_{x}^2+m_{y}^2) U_{pd\pi} - \frac{1}{2}m_{z}(m_{x}^2+m_{y}^2-2m_{z}^2) U_{pd\sigma} \\
U^{A/B}_{p_z,x^2}	&= \langle p^{A/B}_{z,\sigma_s}|\hat{H}_{_{TB}}|d_{x^2-y^2,s}\rangle = \frac{\sqrt{3}}{2} m_{z}(m_{x}^2-m_{y}^2) U_{pd\sigma} - m_{z}(m_{x}^2 - m_{y}^2) U_{pd\pi} \\
U^{A/B}_{p_z,\{x,y\}z}	&= \langle p^{A/B}_{z,s}|\hat{H}_{_{TB}}|d_{\{x,y\}z,s}\rangle = \sqrt{3} m_{z}^2 m_{\{x,y\}} U_{pd\sigma}+(1-2m_{z}^2) m_{\{x,y\}} U_{pd\pi} \\
U^{A/B}_{p_z,xy}	&= \langle p^{A/B}_{z,s}|\hat{H}_{_{TB}}|d_{xy,s}\rangle = m_{z} m_{x} m_{y}(\sqrt{3} U_{pd\sigma} - 2U_{pd\pi}).
\label{us}
\end{split}
\end{equation}
\end{widetext}

\noindent where, for simplicity we have droped the site $j$-index, and the components of the unit vector $\bm m= (m_x,m_y,m_z)$,  are $m_{l}=\{{\nu_l,\mu_l}\}$ specified by the cosine directors between the carbon atoms  at $\{A,B\}$ to the nearby nickel(cobalt) atom, respectively,  with  $l=\{x,y,z\}$. Note that in the $AC$ stacking configuration chosen here, the overlapping between the graphene atoms of sublattice $A$ and  Ni(Co)  are all zero from symmetry ($\nu_x=\nu_y=0$, and $\nu_z=-1$), with the sole exception of the atomic  $p_z$  orbitals  coupling  the $d_{3z^2-r^2}$-orbitals,  namely $U^{A}_{p_z,z^2}\equiv U_{pd\sigma}$.  For carbon atoms in sublattice $B$  we have $U^{B}_{p_z,xz}= \mu_{jx} \widetilde{U}^{B}_{p_z,xz}$, $U^{B}_{p_z,yz}= \mu_{jy} \widetilde{U}^{B}_{p_z,xz}$, $U^{B}_{p_z,xy}=\mu_{jx}\mu_{jy}\widetilde{U}^{B}_{p_z,xy}$, and $U^{B}_{p_z,x^2}=\frac{1}{2}(\mu^2_{jx}-\mu^2_{jy})\widetilde{U}^{B}_{p_z,xy}$. With the matrix elements  
\begin{eqnarray}
\widetilde{U}^{B}_{p_z,xz} & = & \sqrt{3} \mu_{z}^2 U_{pd\sigma}+(1-2\mu_{z}^2) U_{pd\pi} \\\widetilde{U}^{B}_{p_z,xy} & = & \mu_{z}(\sqrt{3} U_{pd\sigma} - 2U_{pd\pi}). 
\end{eqnarray}

\noindent where $U_{pd\sigma}$ and $U_{pd\pi}$ are the Slater-Koster parameters corresponding to overlaps between graphene $p_z$ and nickel(cobalt) $d$-orbitals that can be fitted to first principal calculations. 
\\

 \subsection{Atomic spin-orbit interaction}
 
 The dominant intrinsic spin-orbit interaction in the graphene/Ni(Co) interface will arise among the outer shell $d$-orbitals of nickel(cobalt). Spin-orbit interaction of carbon atoms and between different atoms are assumed to be negligible. The local spin-orbit interaction is modeled through the term,\cite{Konschuh, Min}
\begin{equation}
\label{H}
H_{so}=\frac{\hbar}{2m^2c^2}\left(\nabla V( \bm{r})\times \bm{p} \right)\cdot\bm{s}=\xi ( r )\bm{L} \cdot\bm{S},
\end{equation}
where $m$ is the free electron mass, $c$ is the speed of light, $V( \mathbf{r})$ is the potential of the nickel(cobalt) ions, $ \bm{p}$ is the linear momentum operator, $\bm{s}=(s_{x},s_{y},s_{z})$  with $s_i$,  $i=\{x,y,z\}$ the standard Pauli matrices, $\bm{S}=\frac{\hbar}{2}\bm{s}$ is the electron spin vector operator, and  $\bm{L}$ is the orbital angular momentum operator.  Within the two center approximation the atomic potential is deem to be spherical symmetric,  $V( \bm{r})\rightarrow V( r) $ , then $\nabla V(r)=\hat r dV/dr.$ Hence the function 
\begin{equation}
\xi(r)=\frac{\hbar}{2m^2c^2}\frac{1}{r}\frac{dV}{dr}
\label{xi}
\end{equation}

\noindent contains all the radial dependence.  We are interested in the the matrix elements 
\begin{equation}
\langle \phi^{Ni(Co)}_{\nu,s}|{H_{so}}|\phi^{Ni(Co)}_{\nu',s'}\rangle = \langle \phi^{Ni(Co)}_{\nu,s}|\xi ( r )\bm{L} \cdot\bm{S}|\phi^{Ni(Co)}_{\nu',s'}\rangle
\end{equation}

\noindent  by writing $\{ |\phi^{Ni(Co)}_{\nu,s}\rangle \} $ in the basis of the eigenkets of the angular momentum $\{| l,m \rangle\}$, with $l=2$ and $m=0,\pm1,\pm2$ spanning all the $3d$-orbitals of nickel(cobalt), matrix elements of the form $\xi_l\langle l,m |\bm{L} \cdot\bm{S}|l',m'\rangle \delta_{l,l'}$ has to be calculated in which  $\xi_{l}=\int_{0}^{\infty} \xi(r) R_l^2(r) r^2 dr$ is the  intrinsic spin-orbit coupling parameter, where  $R_l$ is the unknown radial wave function for nickel(cobalt) atoms. For $l=2$ we use the notation  $\xi_{l=2}\rightarrow\xi_{d}$. Because the explicit form of Eq. (\ref{xi}) neither $R_l(r)$ are known, the value of $\xi_{d}$ has to be determined from DFT calculations. In this paper we use $\xi_d = 0.11$\,eV for nickel and $\xi_d = 0.04$\,eV for cobalt, which are similar to the values presented in Table IV.\cite{Barreteau}

Finally the full $10\times10$ matrix that describes the spin-orbit interaction in the extended basis $\{d_{xz,\uparrow},d_{xz,\downarrow},d_{yz,\uparrow},d_{yz,\downarrow},d_{3z^2-r^2,\uparrow},d_{3z^2-r^2,\downarrow},d_{xy,\uparrow},d_{xy,\downarrow}, \\
d_{x^2-y^2,\uparrow},d_{x^2-y^2,\downarrow}\}$ is given by,
\begin{widetext}
\begin{equation}
H_{so}^{Ni(Co)} =\left(\begin{array}{c c c c c c c c c c}
0      		& 0 			&   -i \xi_{d}            	&  0    			&  0				& \sqrt{3} \xi_{d}	& 0 			& i \xi_{d}		& 0 			& -\xi_{d}	\\
0      		& 0	&   0			      	&  i \xi_{d}  		&  -\sqrt{3} \xi_{d}	& 0				&  i \xi_{d} 	& 0			& \xi_{d} 		& 0	\\
 i \xi_{d}      		& 0				&  0         	&  0   		&  0				& -i \sqrt{3} \xi_{d}	& 0 			& - \xi_{d}		& 0 			& -i \xi_{d} \\
   0      			&  -i \xi_{d} 		& 0	      	&  0   	&  -i \sqrt{3} \xi_{d}	& 0				& \xi_{d} 		& 0			& -i \xi_{d}		& 0	\\
   0      			& -\sqrt{3} \xi_{d}	&  0            		& i \sqrt{3} \xi_{d}    	&0 & 0		& 0 			& 0			& 0 			& 0	\\
   \sqrt{3} \xi_{d}      	& 0 				& i \sqrt{3} \xi_{d}	&  0    			& 0		&0 & 0 		& 0			& 0 			& 0	\\
   0      			&  -i \xi_{d} 		&  0            		&  \xi_{d}    		&  0				& 0				& 0	& 0		& 2 i\xi_{d} 	& 0	\\
    -i \xi_{d}      		&  0 				& - \xi_{d}            	&  0    			&  0				& 0				& 0	&0	& 0 			& -2 i\xi_{d} \\
   0      			&  \xi_{d}			&  0            		&  i \xi_{d}     		&  0				& 0				& -2 i\xi_{d} 	& 0			&0 	& 0 \\ 
   -\xi_{d}      		&  0				&  i \xi_{d}            	&  0    			&  0				& 0				& 0 			& 2 i\xi_{d}	& 0	& 0\\		   
\end{array}\right).
\label{hso}
\end{equation}
\end{widetext}

 \subsection{Exchange field coupling}

When graphene is put in proximity with a ferromagnetic layer its electrons experiences an exchange field that leads to the breaking the time reversal symmetry of its bandstructure. Such induced exchange field occurs due to quantum virtual hopping between the graphene layer and the ferromagnetic atoms, removing the degeneracy of the spins states. Thus for collinear ferromagnetic states, spin electrons with up/down orientations (majority/minority spin bands) will experience different energies depending whether their spin is parallel/antiparallel to the local magnetic moment of the ferromagnetic atoms.  In what follows we will adopt a simple tight-binding Stoner Hamiltonian \cite{Stoner} that models the electron spin splitting through a local (on-site) potential energy.   

The  tight-binding Stoner model depends on the local magnetization $\bm{M}_{i,d}=$ of atom $i$ summed over the orbitals of character $d$ since in transition metals the magnetization is mainly carried out by the $3d$-orbitals.  In spin-space the Stoner potential has the form \cite{Barreteau}
\begin{equation}
\bm{V}_{i}^{Stoner}=-\frac{1}{2}I_{i,d}(\bm{M}_{i,d}\cdot\bm{s})
\label{stoner}
\end{equation}
\noindent where  $I_{i}$ is the Stoner parameter for the atom $i$, and $\bm{s} $ is the vector of the spin Pauli matrices.  In this work we chose the magnetization the ferromagnetic nickel(cobalt) layer to be along the $y$-axis, $\bm{M}=M\hat{y}$, thus the Hamiltonian describing the induced exchange field arisen due to the $d$-orbitals of the ferromagnetic layer in the spin-space is parametrized as follows
\begin{equation}
{\cal{H}}_{ex}=\Delta_{ex}s_y
\end{equation}

\noindent where the exchange splitting energy $\Delta_{ex}=|2\Delta_o| $ with $\Delta_o=-\frac{1}{2}I_{i,d} M_{i,d}$, in particular the splitting energy $\Delta_o=0.25$\,eV for nickel and 
 $\Delta_o=0.8$\,eV for cobalt atoms.\cite{Barreteau}

\begin{widetext}
\begin{equation}
{\cal H}_{ex} =
\left(\begin{array}{c c c c c c c c c c}
 0 	      		& -i \Delta_{ex}			&  0         	&  0    			&  0				&0	& 0 			& 0	& 0 			&0	\\
 i\Delta_{ex}      		& 0			&   0			      	& 0		& 0	& 0				& 0 	& 0			&0 		& 0	\\
0      		& 0				&0    	        	&  -i\Delta_{ex}   		&  0				& 0	& 0 			& 0		& 0 			& 0 \\
   0      			& 0 		&  i\Delta_{ex}		      	&0  	   	& 0	& 0				&0		& 0			&0	& 0	\\
   0      			&0	&  0            		&0    	&0	&-i\Delta_{ex}		& 0 			& 0			& 0 			& 0	\\
    	& 0 				&0	&  0    			&  i\Delta_{ex}			&0	& 0 		& 0			& 0 			& 0	\\
   0      			& 0		&  0            		&0 		&  0				& 0				&0 		& -i\Delta_{ex}		&  	& 0	\\
 0		&  0 				&0         	&  0    			&  0				& 0				& i\Delta_{ex}		& 0    		& 0 			& 0 \\
   0      			&0		&  0            		& 0	&  0				& 0				& 0	& 0			&0    	 	& -i\Delta_{ex} \\ 
0    		&  0				&0           	&  0    			&  0				& 0				& 0 			& 0	& i\Delta_{ex} 		&0  	 \\		   
\end{array}\right),
\label{hchi}
\end{equation}
\end{widetext}

\subsection{Full tight-binding Hamiltonian }

Taking into account the spin-orbit and exchange interactions in the full Hamiltonian for graphene/Ni(Co) interface we find

 \begin{equation}
 \cal H =
 \left(\begin{array}{ c  c }
  \cal H_\pi					& {\cal U} \\ 
   {\cal U}^{\dagger}  	 		&\cal H_{{\chi}}  \\   		   
\end{array}\right) ,
\label{Ht}
\end{equation}

\noindent where
\begin{equation}
{\cal H}_\pi =H_\pi \otimes \{\uparrow,\downarrow\}
\end{equation}

 \noindent  specifies the freestanding graphene $\pi$-bands, the graphene-nickel(cobalt) coupling part is given by 
\begin{equation}
  {\cal U} =U \otimes \{\uparrow,\downarrow\},  
  \end{equation}
 
 \noindent and  ${\cal H}_{{\chi}} = E_{Ni(Co)}+H_{so}^{Ni(Co)} + H_{ex}$, with $E_{Ni(Co)}={\cal E}_{Ni(Co)}\otimes \{\uparrow,\downarrow\}$ incorporates the one-site, spin-orbit and exchange field contribution of the Ni(Co) atoms.
Explicitly, In the extended basis $\{d_{xz,\uparrow},d_{xz,\downarrow},d_{yz,\uparrow},d_{yz,\downarrow},d_{3z^2-r^2,\uparrow},d_{3z^2-r^2,\downarrow},d_{xy,\uparrow},d_{xy,\downarrow}, \\
d_{x^2-y^2,\uparrow},d_{x^2-y^2,\downarrow}\}$,  the matrix ${\cal H}_{{\chi}}$ has the form,
 
 \begin{widetext}
\begin{equation}
{\cal H}_{{\chi}} =
\left(\begin{array}{c c c c c c c c c c}
 \varepsilon_{d_1} -  \varepsilon_p    	      		& -i \Delta_{ex}	 			&   -i \xi_{d}            	&  0    			&  0				& \sqrt{3} \xi_{d}	& 0 			& i \xi_{d}		& 0 			& -\xi_{d}	\\
 i\Delta_{ex}	      		& \varepsilon_{d_1} -  \varepsilon_p    			&   0			      	&  i \xi_{d}  		&  -\sqrt{3} \xi_{d}	& 0				&  i \xi_{d} 	& 0			& \xi_{d} 		& 0	\\
 i \xi_{d}      		& 0				& \varepsilon_{d_1} -  \varepsilon_p    	        	&  -i\Delta_{ex}	    		&  0				& -i \sqrt{3} \xi_{d}	& 0 			& - \xi_{d}		& 0 			& -i \xi_{d} \\
   0      			&  -i \xi_{d} 		&  i\Delta_{ex}			      	& \varepsilon_{d_1} -  \varepsilon_p    	   	&  -i \sqrt{3} \xi_{d}	& 0				& \xi_{d} 		& 0			& -i \xi_{d}		& 0	\\
   0      			& -\sqrt{3} \xi_{d}	&  0            		& i \sqrt{3} \xi_{d}    	&\varepsilon_{d_o} -  \varepsilon_p    	&-i\Delta_{ex}			& 0 			& 0			& 0 			& 0	\\
   \sqrt{3} \xi_{d}      	& 0 				& i \sqrt{3} \xi_{d}	&  0    			&  i\Delta_{ex}				&\varepsilon_{d_o} -  \varepsilon_p    	& 0 		& 0			& 0 			& 0	\\
   0      			&  -i \xi_{d} 		&  0            		&  \xi_{d}    		&  0				& 0				&\varepsilon_{d_2} -  \varepsilon_p    		& -i\Delta_{ex}			& 2 i\xi_{d} 	& 0	\\
    -i \xi_{d}      		&  0 				& - \xi_{d}            	&  0    			&  0				& 0				& i\Delta_{ex}	 		& \varepsilon_{d_2} -  \varepsilon_p    		& 0 			& -2 i\xi_{d} \\
   0      			&  \xi_{d}			&  0            		&  i \xi_{d}     		&  0				& 0				& -2 i\xi_{d} 	& 0			&\varepsilon_{d_2} -  \varepsilon_p    	 	& -i\Delta_{ex}	 \\ 
   -\xi_{d}      		&  0				&  i \xi_{d}            	&  0    			&  0				& 0				& 0 			& 2 i\xi_{d}	& i\Delta_{ex}	 		&\varepsilon_{d_2} -  \varepsilon_p    	 \\		   
\end{array}\right).
\label{hchi}
\end{equation}
\end{widetext}

\begin{widetext}
\section{Effective Low-Energy Hamiltonian }

We use the standard band folding method to find the low-energy Hamiltonian of graphene/Ni(Co) system and focus our attention to the modifications of the $\pi$-bands associated to $ \cal H_\pi$ due the rest of the interactions.  For this we require that eigenvalues  $\langle {\cal H}_{{\chi}}  \rangle\ll\langle {\cal H}_\pi \rangle$, and that the characteristic energies $\langle {\cal U} \rangle  \ll ( \langle {\cal H}_{{\chi}}  \rangle - \langle {\cal H}_\pi \rangle )$. The eigenvalue equation for  the full Hamiltonian Eq.~(\ref{Ht}) is
\begin{equation}
\left(\begin{array}{ c  c }
  \cal H_\pi					& {\cal U} \\ 
   {\cal U}^{\dagger}  	 		&\cal H_{{\chi}}  \\   		   
\end{array}\right)
\left (\begin{array}{cc} G \\
\chi
\end{array}\right)= E\left (\begin{array}{cc} G \\
\chi
\end{array}\right),
\label{BF1}
\end{equation}

\noindent where  $G=\{  \psi^{A}_{p_{z\uparrow}},\psi^{A}_{p_{z\downarrow}}\psi^{B}_{p_{z\uparrow}}, \psi^{B}_{p_{z\downarrow}}    \}$,
 and 
 $\chi=\{\psi_{d_{xz,\uparrow}},\psi_{d_{xz,\downarrow}},\psi_{d_{yz,\uparrow}},\psi_{d_{yz,\downarrow}},\psi_{d_{3z^2-r^2,\uparrow}},\psi_{d_{3z^2-r^2,\downarrow}},\psi_{d_{xy,\uparrow}},\psi_{d_{xy,\downarrow}},  \psi_{d_{x^2-y^2,\uparrow}},\\
\psi_{d_{x^2-y^2,\downarrow}\}}$ are the wave functions in the graphene and nickel(cobalt) subspace, respectively. The elimination of $\chi$ from Eq.~(\ref{BF1}) gives:
\begin{equation}
\left[{ \cal H}_{\pi}+{\cal U}\left(E-{\cal H}_{\chi}\right)^{-1}{\cal U}^{\dagger}\right ]G=E\, G,
\label{BF2}
\end{equation}
\noindent and expanding $\left(E-{\cal H}_{\chi}\right)^{-1}$ up to first order in $E$, we obtain
$
\left[{ \cal H}_{\pi}-{\cal U} {\cal H}^{-1}_{\chi}{\cal U}^{\dagger}\right]G \approx E\, {\cal S}\, G,
$ where ${\cal S}=1+{\cal U} {\cal H}^{-2}_{\chi}{\cal U}^{\dagger}$. By defining $\Phi={\cal S}^{1/2}G$, normalized  to $|\Phi|^2\approx G^{\dagger}G+\chi^{\dagger}\chi$
to the same order as the new effective Hamiltonian, we get

\begin{equation}
{\cal S}^{-1/2}\left[ {\cal H}_{\pi}-{\cal U H}^{-1}_{\chi}{\cal U}^{\dagger}\right ]{\cal S}^{-1/2}\Phi={\cal H}_{\it eff}\Phi\simeq E\Phi.
\label{BF4}
\end{equation}
Therefore, after using Eqs. (A17),(A18) and (A19); and going to the reciprocal space by using the cosine directors of Figure~\ref{reciprocalvec}, the effective low-energy Hamiltonian for graphene perturbed by its interaction with nickel(cobalt) has the form (assuming ${\cal S}\approx 1$)

\begin{equation}
{\cal H}_{\it eff} ( {\bm p} )\simeq {\cal H}_{\pi}-{\cal U H}^{-1}_{\chi}{\cal U}^{\dagger} =  \left(\begin{array}{c c c c}
  \varepsilon_{_+} + \varepsilon_{_-}    	& 0		&   v^{*}_{_F} p_{_{-}}									                          &  i v^{*}_{_d} p_{_{-}}\\
0				&  \varepsilon_{_+} +  \varepsilon_{_-}   	&  -i v^{*}_{_d}  p_{_{-}}					                          & v^{*}_{_F} p_{_{-}} \\
    v^{*}_{_F} p_{_{+}}						&  i v^{*}_{_d}  p_{_{+}}		& \varepsilon_{_-} -  \varepsilon_{_+} 	         &0 \\
   - i v^{*}_{_d} p_{_{+}}		 		                &  v^{*}_{_F} p_{_{+}}							& 0                                          &  \varepsilon_{_-} -  \varepsilon_{_+} 
\end{array}\right)+
\left(\begin{array}{c c c c}
 0    	& -i \Sigma_{_{+}} 			&  0								                          &  0 \\
  i \Sigma_{_{+}} 				&  0   	&    - i \lambda_{_{\it XR}} 				                          & 0 \\
    0						&   i \lambda_{_{\it XR}} 		& 0 	         &  -i \Sigma_{_{-}} \\
  0	 		                & 0					&  i \Sigma_{_{-}}	                                          &  0
\end{array}\right),
\label{heffpz}
\end{equation}

\noindent where  ${\bm p}=(p_x,p_y)$  is the momentum operator,  with $p_{_{\pm}}=p_{x} \pm ip_{y}$,  and $ v^{*}_{_F}$ and  $v^{*}_{d}$ are renormalized Fermi velocities (see text for their explicit dependence in terms ofthe Slater-Koster parameters). The diagonal parameters $\varepsilon_{\pm}$ are shift energies defined in Eq.(6). The nondiagonal parameters $\Sigma_{\pm}=\Delta_{-}\pm\Delta_{+}$,  with $\Delta_{\pm}$ defined in Eq.(5). Finally the effective spin-orbit parameter $ \lambda_{_{\it XR}} $ is described in Eq.(3). All these parameters are characterized  in a nontrivial way on the overlapping $p_z$-$d$ integrals and on the exchange field. The effective Hamiltonian (\ref{heffpz}) can be written in compact form as given by Eq.(\ref{heffpz1}).
\end{widetext}


\begin{thebibliography}{11}
\bibitem{shortnotation} Hereafter we shal use the short notation $z^2$ for the orbitals $3z^2-r^2$ of the ferromagnetic metals.
\end{thebibliography}
\end{document}